\documentclass[prl,twocolumn, showpacs,superscriptaddress]{revtex4}
\usepackage[pdftex]{graphicx}
\usepackage{amsmath}
\usepackage{amssymb}
\begin{document}

\title{s-wave superconductivity in non-centrosymmetric Re$_3$W probed by magnetic penetration depth}
\author{Yuri L. Zuev}
\affiliation{Material Science and Technology Division, Oak Ridge National Laboratory, Oak Ridge, TN, 37831, USA}
\author{Valentina A. Kuznetsova}
\affiliation{Dept. of Physics, University of Tennessee, Knoxville, TN, 37996, USA}
\author{Ruslan Prozorov}
\affiliation{Ames Laboratory and Department of Physics and Astronomy, Iowa State
University, Ames, IA, 50011, USA}
\author{Matthew D. Vannette}
\affiliation{Ames Laboratory and Department of Physics and Astronomy, Iowa State
University, Ames, IA, 50011, USA}
\author{Maxim V. Lobanov}
\affiliation{Department of Chemistry and Chemical Biology, Rutgers, The State University of New Jersey, Piscataway, New Jersey 08854}
\author{David K. Christen}
\affiliation{Material Science and Technology Division, Oak Ridge National Laboratory, Oak Ridge, TN, 37831, USA}
\author{James R. Thompson}
\affiliation{Material Science and Technology Division, Oak Ridge National Laboratory, Oak Ridge, TN, 37831, USA}
\affiliation{Dept. of Physics, University of Tennessee, Knoxville, TN, 37996, USA}
\begin{abstract}
We report measurements of the temperature dependence of the magnetic penetration depth $\lambda(T)$ in non-centrosymmetric superconductor Re$_3$W. We employed two experimental techniques: extraction of $\lambda(T)$ from  magnetic {\em dc}-susceptibility, measured on a powder sample, and the rf tunnel diode resonator technique, where a bulk polycrystalline sample was used. The results of both techniques agree: the temperature dependence of the penetration depth can be well described by weak-coupling, dirty-limit, s-wave BCS theory where we obtain $\Delta(0)/k_BT_C=1.76$. No evidence for unconventional pairing resulting from the absence of the inversion symmetry is found.
\end{abstract}
\maketitle

Superconductors possessing a crystal structure without an inversion center are a focus of current research~\cite{SZB, Bonalde2005, Yuan, Hayashi, Sergienko, Izawa, Yip, Young, Kimura, Sugitani, Mineev, Frigeri, Fujimoto2007, Bonalde2007}. The lack of inversion symmetry means that parity $P$ is not a good quantum number, i.e. electronic states cannot be labeled as either even or odd under inversion ${\mathbf r}\rightarrow -{\mathbf r}$. As a consequence, a Cooper pair's internal angular momentum ${\mathbf S}$ is not necessarily even ($S=2n$) or odd ($S=2n+1$) under such transformation ($n=$integer). The physical interaction that breaks inversion symmetry is the antisymmetric spin-orbit interaction. Its presence may lead to a pairing state with a mixed singlet-triplet character. This in turn may form nodes in the gap function, easily detectable by  measurements of the magnetic penetration depth $\lambda(T)$~\cite{prozorov2006}.

A fully gapped, isotropic pair\-ing state pro\-du\-ces thermally - activated behavior of the su\-per\-flu\-id density , $\rho_s\propto \lambda^{-2}(T)\propto 1-(2\pi\Delta(0)/T)^{1/2}e^{\Delta(0)/k_BT}$ at low temperatures, meaning that $\lambda^{-2}(T)$ hardly changes at $T<0.3T_C$(for a review see~\cite{prozorov2006}). On the other hand, it is known that line nodes in the gap cause the superfluid density to display a power law behavior, $\lambda^{-2}(T)\propto 1-aT^n$, where $n$ may equal 1 (such is the case in clean $d$-wave cuprates[~\cite{AnnettGoldenfeld, Hardy}]), 2 (dirty cuprates~\cite{GoldenfeldHirschfeld}), 3 ($T^3$ behavior was reported in electron-doped Pr$_{1.86}$Ce$_{0.14}$CuO$_4$~\cite{Skinta} and in certain organic superconductors~\cite{Prozorov2001}). 

As for the compounds without inversion symmetry, a nonexponential temperature dependence of $\lambda^{-2}(T)$ that implies a superconducting gap with nodes has indeed been found in CePt$_3$Si~\cite{Bonalde2005} and Li$_2$Pt$_3$B~\cite{Yuan}.

The model of N. Hayashi \textit{et al.}~\cite{Hayashi} allows  calculation of the temperature dependent $\lambda(T)$ for a given magnitude of singlet and triplet (or $s$-wave and $p$-wave) components. It produces a good agreement with experiment in the case of non-centrosymmetric Li$_2$Pt$_3$B and Li$_2$Pd$_3$B~\cite{Yuan}. 

Superconductivity in the in\-ter\-me\-tal\-lic com\-po\-und Re$_3$W was first studied in the 1960's~\cite{Hulm, Blaugher}. It was found to have a  su\-per\-con\-duc\-ting transition temperature $T_C \sim 9$ K and a crystal structure without inversion center. This makes it a good system for the present study, because we can access the low - temperature region, $T<0.3T_C$ needed to relate change of the penetration depth to the gap structure. Another reason why this material is a good candidate is that atomic numbers $Z$ of both constituents are large, 75 for Re and 74 for W. Large $Z$ promotes spin-orbit interaction, which breaks the inversion symmetry. Re$_3$W crystallizes in a so-called $\alpha$-Mn, or A12 structure. As far as we know, there has been no detailed calculation of electronic spectra for this particular structure or subsequent experimental work on this material since~\cite{Blaugher}. In fact we are aware of only two investigations on Re-W system: a purely metallurgical study by S. Tournier {\em et al.}~\cite{Tournier} and a calculation of stability of various Re-W phases by K. Perrson {\em et al.}~\cite{Perrson}.

\begin{figure}[t]
\centering
\includegraphics[width=\columnwidth]{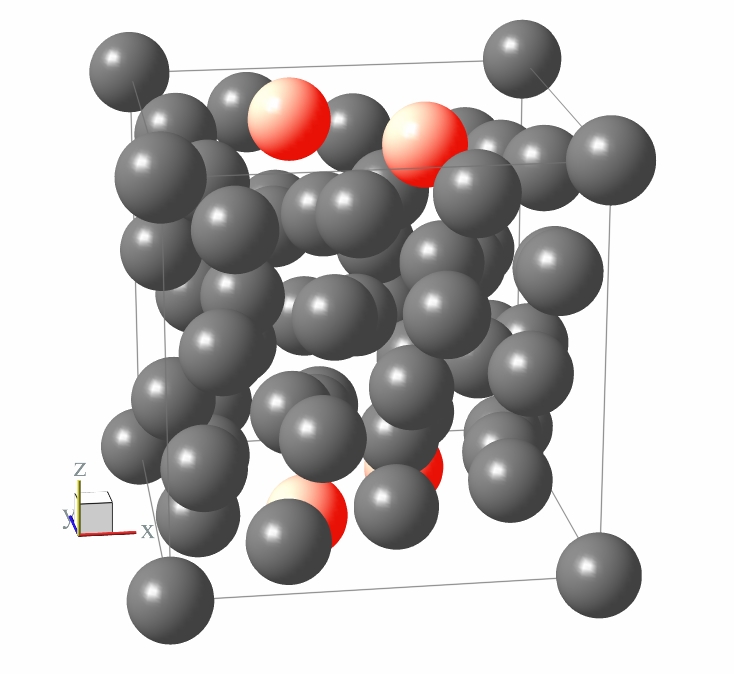}
\caption{(Color online) Crystal structure of non-centrosymmetric Re$_3$W, obtained by Rietveld refinement of powder diffraction data. Rhenium and tungsten atoms cannot be distinguished due to small difference in atomic numbers. Pairs of atoms, shown in red, on top and bottom faces of the cube demonstrate the absence of inversion symmetry.}
\label{fig:struct}
\end{figure}

Re$_3$W samples used in the present study were prepared from elemental starting materials (Alfa AESAR) containing less than 0.001\% total impurities. Powdered Re was mixed with powdered W in molar ratio 3:1, and the mixture was pressed into pellets using a hydraulic press. Subsequently, pellets were individually arc-melted in high-purity Ar atmosphere and then annealed for 14 days at 1773 K in either an atmosphere of flowing high-purity argon, or high vacuum,  and then allowed to cool. Synchrotron X-ray powder diffraction patterns were collected at the X7A 
beamline at the Brookhaven National Laboratory. The diffraction pattern of Re$_3$W can be almost completely indexed assuming $I$-centered cubic ($\alpha$-Mn) lattice (trace amount of Re metal could be detected with estimated weight fraction of
~1\%). The Rietveld refinement showed that studied material belongs to the symmetry group I$\bar{4}$3m, with cubic unit cell size $a = 9.59656(5)$\AA. The symmetry group I$\bar{4}$3m includes an inversion axis of fourth order and therefore lacks inversion center. The crystal structure is shown in Fig.~\ref{fig:struct}. Since X-ray scattering intensity is proportional to the square of the atomic number $Z$, and Re and W differ in $Z$ by 1, they could not be distinguished.  Absence of inversion can be most clearly seen by examining e.g. two pairs of atoms shown in red on the top and bottom faces of the cube. 

We have measured the magnetic penetration depth $\lambda(T)$ by two well-established and independent methods: through measurement of the dc-susceptibility  $\chi(T)$ of a collection of small particles dispersed in epoxy, and from the change in resonance frequency of a tunnel-diode driven oscillator operating at 10 MHz. 

In the first technique, a bulk sample of annealed Re$_3$W ($T_C=7.4$ K) was ball -milled into a powder with resulting particle size of the order of 10 $\mu$m. The particle size distribution and an average aspect ratio of 0.6 were determined by direct optical microscopy analysis. Eight milligrams of powder ($\pm$ 0.1mg) were mixed with epoxy in a gelatin capsule and cured for an hour at room temperature in an applied magnetic field of 6 T. Curing in field induces alignment of the paramagnetic particles with their long dimension parallel to field, thereby reducing their demagnetizing ratio.  Measurement of the \emph{dc} susceptibility of this sample was performed in a Quantum Design MPMS SQUID magnetometer in the applied field of 3 Oersted at temperatures down to 1.85 K (0.25$T_C$).

\begin{figure}[b]
\centering
\includegraphics[width=\columnwidth]{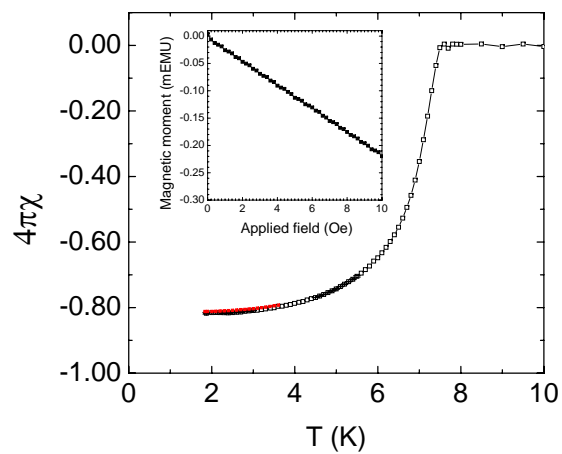}
\caption{(Color online): dc susceptibility vs. $T$ for powder Re$_3$W. Applied field is 3 Oe. Higher resolution data were taken at lower temperature, $T<3.6$K (red). Inset: the magnetic moment is linear in field up to at least 10 Oe ($T=5$ K)}
\label{fig:chi}
\end{figure}

The epoxy and other addenda showed an insignificant amount of diamagnetism, about 0.5\% that of the actual sample in the above temperature range. Care was taken to ensure the linearity of response, i.e. independence of the measured $\chi$ on the applied field strength. The inset to Fig.~\ref{fig:chi} shows that at $T=5$ K the measured magnetic moment is linear in field up to at least 10 Oe. It should be mentioned that samples cured in zero field and likely having randomly oriented particles did not have an appreciable linear region.

For analysis of the experimental data we assume that powder particles have an elongated ellipsoidal shape with short semiaxis $R$ and use the expression
\begin{equation}
4\pi\chi=-\frac{1}{1-D}(1-\frac{3\lambda}{R}\coth\frac{R}{\lambda}+\frac{3\lambda^2}{R^2})
\label{eq:chi}
\end{equation}
to relate the measured susceptibility to $\lambda$. With $D=1/3$ the equation above is appropriate for a sphere, but we use it here with an average particle's  demagnetizing factor $D=0.22$, based on the measured aspect ratio. Next we average over the measured particle size distribution and invert Eq.(\ref{eq:chi}) numerically to obtain $\lambda$. Since it is $\lambda/R$ that enters Eq.(\ref{eq:chi}), it turns out that accurate knowledge of particle sizes is important only for determination of absolute value of $\lambda$, but not for normalized quantity $\lambda(T)/\lambda(0)$. In our case the particles are still large compared with the penetration depth, therefore we do not obtain correct absolute value of $\lambda$. The temperature dependence $\lambda(T)/\lambda(0)$ on the other hand is insensitive to $R$. The absolute value of the penetration depth for our samples was estimated from measurements of the lower and upper critical fields and the Ginzburg-Landau parameter $\kappa$, yielding $\lambda(T=0)=300\pm10$ nm. 

The second technique is the tunnel-diode resonator where a self-resonating LC tank circuit is powered by a tunnel-diode \cite{prozorov2000,prozorov2006}. A bulk polycrystalline sample was used. The sample is inserted into a coil, whose inductance then changes and causes the shift of the resonant frequency, $\Delta f$. This shift is proportional to the magnetic susceptibility, $\chi$, of the sample, thus to the London penetration depth, $\lambda$ (in the Meissner state), $\Delta f=-4\pi\Delta f_{0}\chi$, where $\Delta f_{0}$ is a sample shape and volume dependent calibration constant. At low temperatures, $\chi=-\left(  4\pi\right)  ^{-1}\left(  1-\lambda/R\tanh\left(  R/\lambda\right)  \right)  $ where $R$ is the effective sample dimension \cite{prozorov2000}. High stability (0.1 ppb) and small excitation field amplitude (\symbol{126} 20 mOe) result in sub-Angstrom precision of the measurements.


We will discuss both the  powder sample and bulk sample measurements together. The dc-susceptibility $\chi(T)$ is shown in Fig.~\ref{fig:chi}. Higher resolution data were taken between 1.85 and 3.6 K (shown in red) by averaging the magnetic moment over several measurements at each temperature.

\begin{figure}[t]
\centering
\includegraphics[width=\columnwidth]{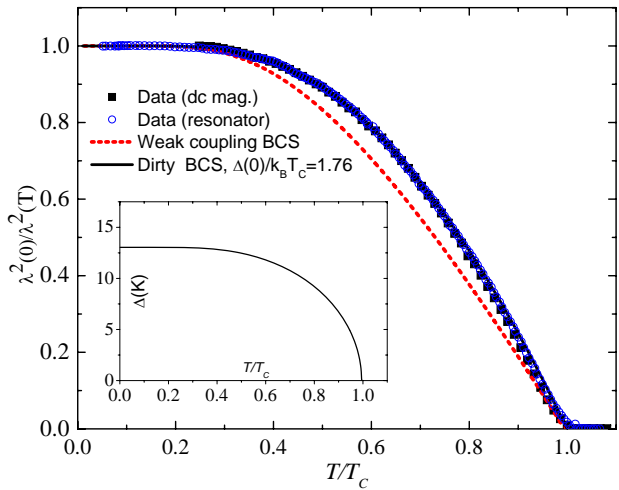}
\caption{(Color online) Temperature dependence of normalized superfluid density $\lambda^2(0)/\lambda^2(T)$.  Squares- data from dc magnetization measurement. Circles- tunnel diode resonator method. Dotted line: clean limit, weak coupling BCS curve $\lambda_L^{-2}(T)$. Solid line: dirty limit BCS calculation with $\Delta(0)/k_BT_C=1.76$, weak coupling value. Inset: energy gap $\Delta(T)$}.
\label{fig:SfD}
\end{figure}

The ex\-trac\-ted normalized su\-per\-flu\-id density $\lambda^2(0)/\lambda^2(T)$ is shown in Fig.~\ref{fig:SfD} by black squares along with the results from the resonator experiment (blue circles). 

Measurements in polycrystals are somewhat difficult to quantify, so we attempt to describe the data from several angles. The main frame, Fig.~\ref{fig:SfD} shows the extracted superfluid density (in the case of the resonator, $\lambda(0)=300$ nm was used in the data analysis). The agreement between the two measurements is excellent. Data are shown by symbols and lines through the data are $s$-wave weak coupling BCS theory in the dirty limit, for which the superfluid density is (theory in~\cite{Tinkham}),
\begin{equation}
\left[\frac{\lambda^2(0)}{\lambda^2(T)}\right]_{\mathrm{dirty}}=\frac{\Delta\left(  T\right) }{ \Delta\left(
0\right)}  \tanh\left(\frac{\Delta\left(  T\right)}{2k_BT}\right)
\label{eq:dirty}
\end{equation}
The temperature dependent gap, $\Delta(T)$, was obtained as a solution of the self-consistent gap equation in the full temperature range (Fig.~\ref{fig:SfD}, inset).
Both data sets can be fit well by the above equation where we obtain $(\Delta(0)/k_BT_C)_{\mathrm{dirty}}=1.76$, the standard BCS value.  It is possible to get a good fit assuming weaker electronic scattering, but this would require slightly increased coupling strength. In order to estimate the maximum deviation from weak coupling, we calculate the London penetration depth in the clean limit according to~\cite{Tinkham}
\begin{equation}
\left[\frac{\lambda^2(0)}{\lambda^2(T)}\right]_{\mathrm{clean}}=1+\int_{\Delta(T)}^{\infty}\frac{\partial f}{\partial E}\frac{E}{\sqrt{E^2-\Delta^2(T)}}\ dE
\label{eq:clean}
\end{equation}

\begin{figure}[b]
\centering
\includegraphics[width=\columnwidth]{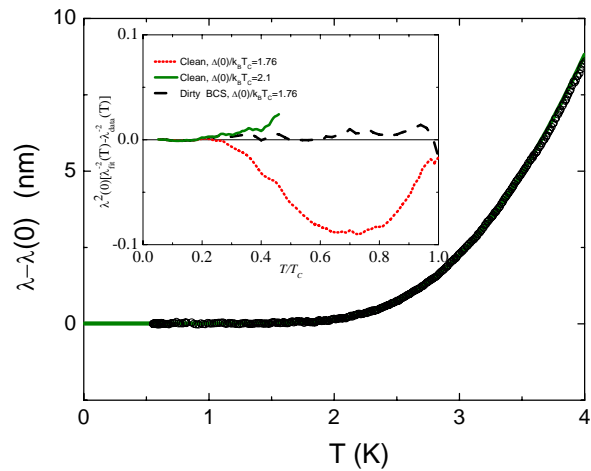}
\caption{(Color online) Low temperature change in $\lambda(T)$ relative to zero-temperature value $\lambda(0)$. Solid green line is a clean limit BCS fit, producing $(\Delta(0)/k_BT_C)_{{\mathrm{clean}}}=2.1$. Inset: quality of fit for dirty limit, Eq.~\ref{eq:dirty}, (black dash) and clean limit  with $(\Delta(0)/k_BT_C)_{{\mathrm{clean}}}=1.76$ (red dots) and 2.1 (green solid line)}  
\label{fig:lambda}
\end{figure}

where $E=[(\varepsilon-\mu)^2+\Delta(T)^2]^{1/2}$ is the elementary excitation energy and $f=[e^{E/k_BT}+1]^{-1}$ is the Fermi function. This expression assumes isotropic  $s$-wave pairing state and spherical Fermi surface. The dashed line in Fig.~\ref{fig:SfD} is the clean limit BCS result with $\Delta(0)=1.76k_BT_C=1.1$ meV, shown for comparison. Loosening the constraint on $\Delta(0)$ and treating it as an adjustable parameter, we performed fitting of the low-temperature portion of the $\lambda(T)$, as shown by the thin solid line in Fig.~\ref{fig:lambda}. The advantage of this procedure is that it does not require assumptions regarding $\lambda(0)$. The best fit was achieved at $(\Delta(0)/k_BT_{C})_{\mathrm{clean}}=2.1\pm0.1$. This serves only as an upper bound on the gap magnitude. As discussed below, we believe this material to be in the dirty limit, where Eq.(\ref{eq:dirty}) applies. The dirty BCS fit to the data is essentially perfect in the entire temperature range, as shown in the inset to figure~\ref{fig:lambda}.



The main conclusions that we draw from the data are that superfluid density in these Re$_3$W samples is adequately described within a BCS framework and no appreciable influence of spin-triplet pairing is detected. This makes Re$_3$W different from a CePt$_3$Si, but similar to Li$_2$Pd$_3$B~\cite{Yuan}. In the latter case, spin-triplet component of the order parameter was noticed, but it was smaller than spin-singlet, thus making for a nodeless  (but anisotropic) gap. In our case however no contribution from spin-triplet is detected at all. This is somewhat surprising, given large atomic numbers of both Re and W, as discussed in the introduction.

A possible reason for the absence of a spin-triplet component may be significant disorder. From resistivity measurements we estimated an electronic mean free path of 1.5 nm, while the BCS coherence length is much longer, $\xi_0=\hbar v_F/\pi\Delta_0\approx 200$ nm with a reasonable assumption about Fermi velocity, $v_F$. Thus the material appears to be quite dirty, as mentioned before. If the order parameter has nodes, even non-magnetic impurities suppress $T_C$ very effectively (this happens for example in high-$T_C$ cuprates). It is possible that triplet component has been effectively suppressed by scattering. In references~\cite{Mineev, Frigeri} the effect of impurities on mixed-pairing superconductivity was considered. The end result was that unconventional pairing channel should be suppressed by non-magnetic impurities similarly to Abrikosov-Gorkov $T_C$ suppression in conventional superconductors by impurities with permanent magnetic moments. In contrast, disorder reduces $T_C$ in the conventional channel, but does not suppress it to zero. 

Finally, in $d$-wave cuprates, scattering fills in electronic states at the gap nodes, thereby suppressing the superfluid density at low temperatures and changing $T$-linear to $T^2$ behavior~\cite{GoldenfeldHirschfeld}. It is possible that a similar mechanism is at play here, masking a power law behavior of $\lambda^{-2}(T)$.  

In closing, we suggest that Re$_3$W material with less disorder should be explored to clarify the existence of mixed-parity pairing in this compound. 
 
We would like to thank Dr. I. A. Sergienko for bringing this material to our attention, V. G. Kogan for useful discussions, D. G. Mandrus for valuable comments on the manuscript. Work at ORNL was sponsored by the division of Materials Sciences and Engineering, Office of Basic Energy Sciences and Office of Electricity Delivery and Energy Reliability, U. S. DOE, under contract DE-AC05-00OR22725 with Oak Ridge National Laboratory, managed by UT-Battelle, LLC. Work at the Ames Laboratory was supported by the Department of Energy-Basic Energy Sciences under Contract No. DE-AC02-07CH11358. YLZ would like to acknowledge support from ORISE. R. P. acknowledges support from the NSF grant number DMR-05-53285 and the Alfred P. Sloan Foundation.

\end{document}